\documentclass[aps,prb,twocolumn,superscriptaddress,showpacs,showkeys]{revtex4}
\usepackage{bm,amsmath,graphicx,dcolumn}
\begin{document}

\title{Automatic physical phasing X-ray crystallography}
\author{S\'ergio L. Morelh\~ao}
\email{morelhao@if.usp.br}
\affiliation{Instituto de F\'{\i}sica, Universidade de S\~ao Paulo, CP 66318, 05315-970 S\~aoPaulo, SP, Brazil}
\author{Luis H. Avanci}
\affiliation{Instituto de F\'{\i}sica, Universidade de S\~ao Paulo, CP 66318, 05315-970 S\~aoPaulo, SP, Brazil}
\author{Stefan Kycia}
\affiliation{Laborat\'orio Nacional de Luz S\'{\i}ncrotron/LNLS, CP 6192, 13084-971 Campinas, SP, Brazil}

\date{\today}
\begin{abstract}
Phase invariants are important pieces of information about the atomic structures of crystals. There are several mathematical methods in X-ray crystallography to estimate phase invariants. The multi-wave diffraction phenomenon offers a unique opportunity of physically measuring phase invariants. In this work, the underneath principals for developing an automatic procedure to extract accurate phase-invariant values are described. A general systematic procedure is demonstrated, in practice, by analyzing intensity data from a KDP crystal.
\end{abstract}



\maketitle

\section{Introduction}
In X-ray crystallography, the phases of the diffracted waves are roughly estimated by mathematical methods, know as {\em Direct Methods}~\cite{[1],[2]}, for analyzing intensity data sets composed of a large number of reflections. These methods exploit algebraic or probabilistic relationships among the phase values. Some of such relationships are triplet phase invariants; they are invariant from the choice of origin in the crystal lattice. Experimental procedures allowing physical measurements of phase invariants are of great interest since, in principle, they could extend the efficiency of the {\em Direct Methods} to complex structures such as proteins. It would have to be compared to other procedures that are actually used to the same purposes, such as multiple anomalous dispersion and multiple isomorphous replacement~\cite{[3]}.
\begin{figure}
\includegraphics[width=2.1in]{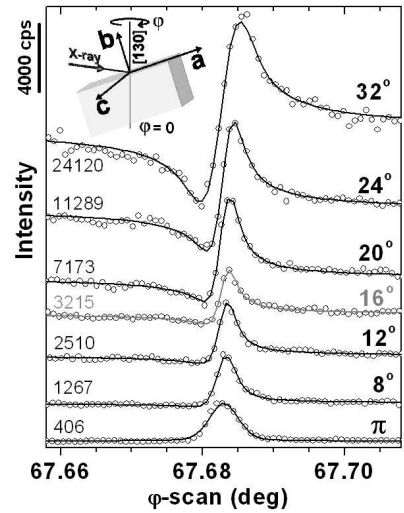}
\caption{\label{fig1} Experimental (open circles) and simulated (solid lines) $\varphi$-scans of the $260/11\bar{2}/152$ three-beam diffraction in a KDP crystal taken at different polarization angle, $\chi$ (right-hand side of each scan). [001] is the reference direction ($\varphi=0$, see inset), X-ray photon energy is 7482eV, and further experimental details can be found elsewhere~\cite{[8]}. The intensity scale is linear, but for visualization purposes the ordinates of some scans are shifted from their actual values, given at the left (in cps). The $\varphi$-scan at $\chi=16^{\circ}$ (gray scan) was mistakenly collected at the shoulder of the 260 reflection ($\Delta\omega=0.003^{\circ}$, 30\% of the FWHM~$=0.01^{\circ}$). The flexibility of the fitting equation, Eq.~(\ref{eq1}), to reproduce these $\varphi$-scans is exploited in Fig.~\ref{fig3}(b).}
\end{figure}

Physical measurements of triplet phase invariants are possible by means of three-beam diffraction (3BD) experiments~\cite{[4],[5]} where the interference of simultaneously diffracted waves provide information on phase values. However, besides all experimental and analytical difficulties involved in phase determination from 3BD experiments, the most serious and practical limitation of physical-phasing crystallography (PPC) is the reduce number of 3BD cases suitable for phasing. The reciprocal space of complex molecule crystals are full of reflections where isolated 3BD cases have become even more rare; phasing general $n$-beam cases ($n>3$) is not feasible at the moment due to theoretical deficiencies. Therefore, it is important to mention that, regarding complex crystals, the usefulness of PPC is quite limited when compared to the available phasing procedures. Nevertheless, there are researches focused on developing and optimizing experimental data collection procedures for PPC~\cite{[6]}. On the other hand, the 3BD experiments offer an unique opportunity for accurate determination of triplet phase invariants, and consequently, for studying crystalline structures via measurements of these invariants. For example, depending on the achieved experimental accuracy, electron density of chemical bonding charges~\cite{[7]} or even distortion of molecules under applied electric field can be investigated by monitoring a few triplet phases. Note that each triplet phase is an absolute value since it already is the phase differences between two diffracted X-ray waves, and not a relative quantity such as obtained in peak position or intensity measurements.

This work has been motivated by our desired of developing at LNLS a systematic and practical procedure for determining phase invariants with good accuracy. Experimental and analytical procedures are still to be improved to push phase measurements from the state-of-art to routinely and automatic phasing procedures; otherwise it will be very difficult to non-expert users to take advantages of the new possibilities offered by measuring this physical quantity. Data collection procedures are already proposed~\cite{[4]}, and undergoing improvement~\cite{[8]}, but the actual challenger is the data analysis procedure~\cite{[5]}. Here, we outline the underneath principals for developing an automatic procedure to extract accurate phase values from 3BD interference profiles. A general systematic procedure is demonstrated, in practice, by analyzing 3BD intensity data from a KDP crystal, and the major sources of errors are pointed out.

\section{Theoretical basis}
\begin{figure*}
\includegraphics[width=3.2in]{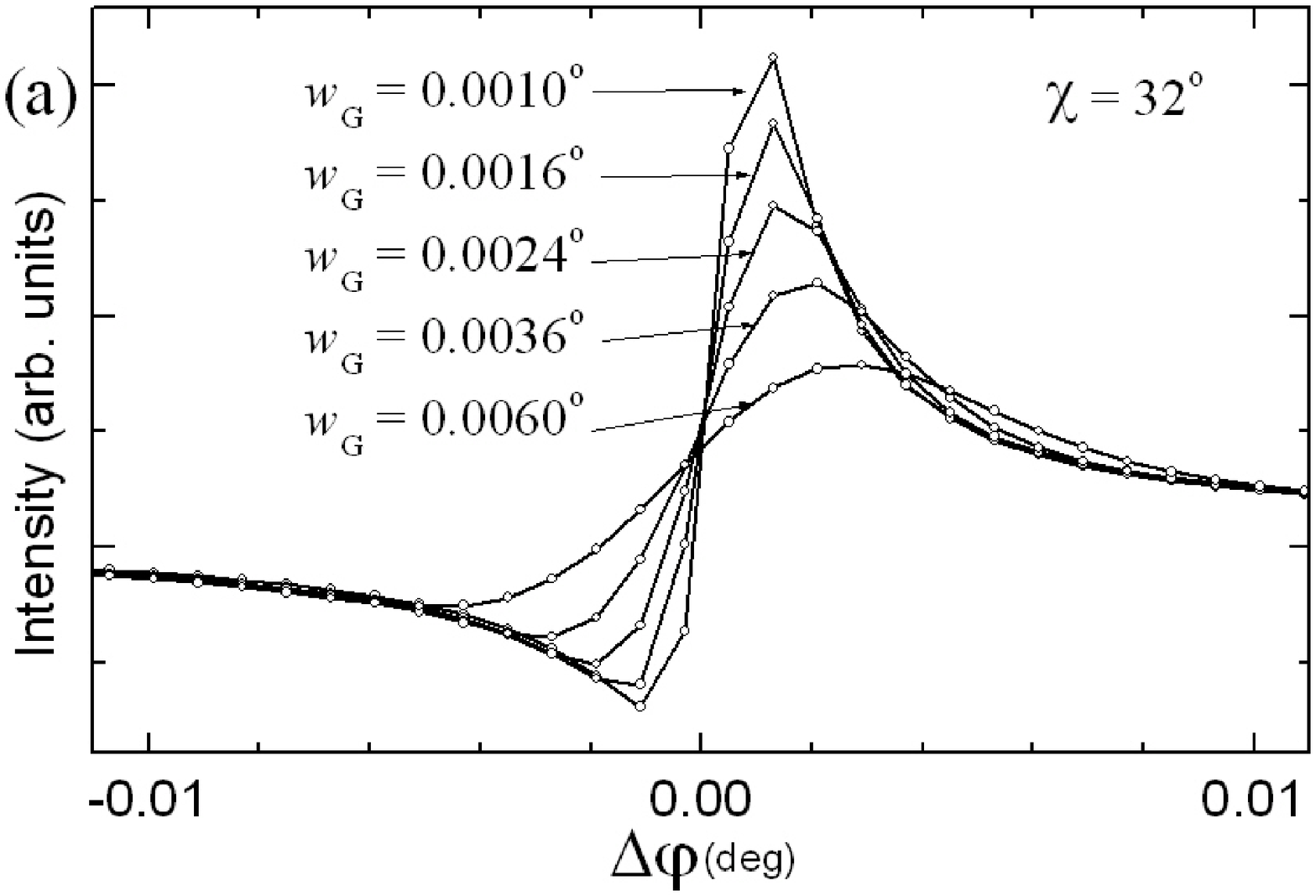}
\includegraphics[width=3.2in]{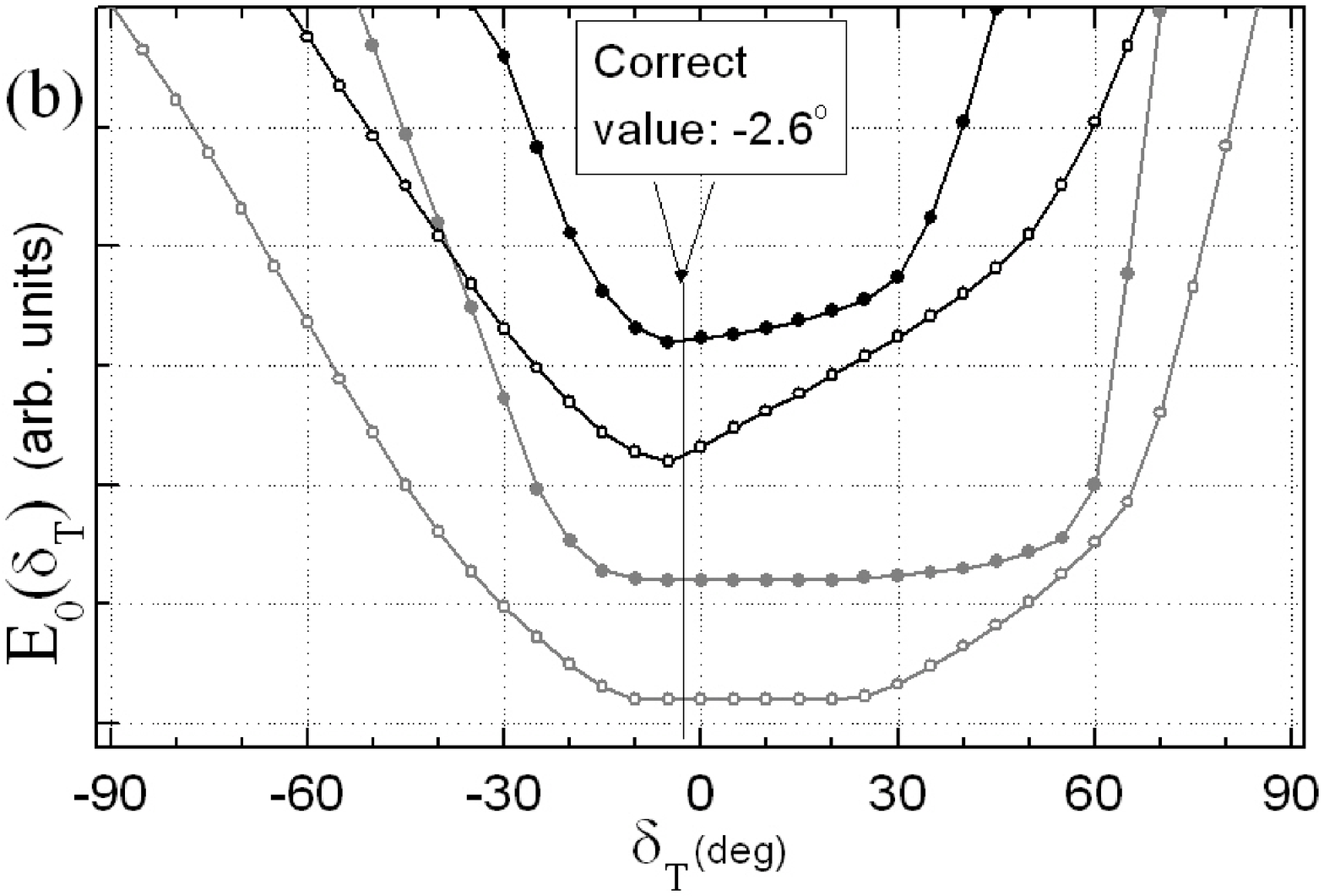}
\caption{\label{fig2} (a) Simulated instrumental broadening effects on $\varphi$-scans. Simulation parameters used into Eq.~(\ref{eq1}): $\delta_{\rm T}=-2.6^{\circ}$, $\chi=32^{\circ}$, and $\bm{p} = [0.0012^{\circ},\> 1.0,\> 0.8,\> 0.0,\> 67.683^{\circ},\> w_{\rm G}]$ where the instrumental width values, $w_{\rm G}$, are indicated by arrows. $\Delta\varphi = \varphi - \varphi_0$. (b) Theoretical accuracy in phase measurements as a function of the instrumental width $w_{\rm G}$, and amplitude ratio $R$. The $E_0(\delta_{\rm T})$ curves were obtained by fitting the profiles in (a) with $w_{\rm G}=0.001^{\circ}$ (open circles) and $w_{\rm G}=0.006^{\circ}$ (closed circles). The fittings have been carried out by the DEA within the allowed ranges: $\bm{p} = [0.0008^{\circ} : 0.0012^{\circ},\> R,\> 0.2:1.0,\> 0.0,\> \varphi_0\pm0.012^{\circ},\> 0.001^{\circ} : 0.007^{\circ}]$ where $R=1.0$ (black lines) or $R=[0.6:1.4]$ (gray lines). Definition on the $\partial E_0 / \partial \delta_{\rm T} = 0$ position gives the accuracy on $\delta_{\rm T}$. }
\end{figure*}

In general, the 3BD intensity profiles are dominated by the interference of two diffracted waves. It leads to a relatively simple parametric equation that can be used to fit most of the experimental intensity profiles and to extract the phase values. It is given by~\cite{[5]}

\begin{widetext}
\begin{equation}
\texttt{I}(\varphi)=\int_{-u_0}^{+u_0}(1-b|\texttt{f}(u)|^2)\left[|\bm{D}_{\rm
A}|^2+|\bm{D}_{\rm BC}(u)|^2+\xi \bm{D}_{\rm A} \cdot \bm{D}_{\rm BC}^{\ast}(u) +\xi \bm{D}_{\rm
A}^{\ast} \cdot \bm{D}_{\rm BC}(u)\right]\texttt{G}(\varphi-u){\rm d}u
\label{eq1}
\end{equation}
\end{widetext}
where $\bm{D}_{\rm A}=\bm{D}_0 \bm{v}_{\rm A}$ and $\bm{D}_{BC}(\varphi)= \bm{D}_0 R \texttt{f}(\varphi){\rm e}^{i\delta_{\rm T}}\bm{v}_{\rm BC}$ are the amplitudes of the primary and secondary electric displacement wavefields generated by the primary reflection, A, and by the detour reflection (also known as {\em Umweg} reflection) formed by two consecutive reflections, B and C. $R$ stands for maximum amplitude ratio of these waves. $\bm{v}_{\rm A}$ and $\bm{v}_{\rm BC}$ are polarization factors for linearly polarized incident radiation. $\delta_{\rm T}$ is the phase difference between these waves, which is the triplet phase invariant. A gaussian convolution, $\texttt{G}(u)$ with ${\rm FWHM}=w_{\rm G}$ and $u_0=\pm2.5w_{\rm G}$, is necessary to account for the instrumental width $w_{\rm G}$. $\texttt{f}(\varphi)=w_{\rm S}/[2(\varphi-\varphi_0)-iw_{\rm S}]$ is a line profile function (FWHM $=w$, $w_{\rm S}=\pm w$) describing the intrinsic 3BD profile as a function of the azimuthal rotation angle $\varphi$. $b$ and $\xi$ are related to energy balance mechanisms among the diffracted beams and crystalline imperfections, respectively~\cite{[5]}.

Essentially, the analytical problem in accurate phase determination resides on how to adjust the vector of parameters, $\bm{p} = [w, R, \xi, b, \varphi_0, w_{\rm G}]$, without compromising the extracted values for $\delta_{\rm T}$. Here, a simple and fast evolutionary algorithm (DEA)~\cite{[9]} has been used for fitting the experimental profiles where the improvements of the fittings are guided by a mean-absolute deviation function, $E(\bm{p})$. The basic strategy is then to find out the minimum of $E(\bm{p})$ as a function of $\delta_{\rm T}$, i.e. $E_0(\delta_{\rm T})$, while {\bf p} is kept within reasonable ranges of allowed values. The minimum of the $E_0(\delta_{\rm T})$ curve, $\partial E_0 / \partial \delta_{\rm T} = 0$, provide the experimental value for $\delta_{\rm T}$.

\section{Results and discussions}

Fig.~\ref{fig1} shows set of 3BD data collected at Brazilian Synchrotron Light Laboratory (LNLS) with the polarimeter-like diffractometer described elsewhere~\cite{[8]}. It is composed of several $\varphi$-scans taken at different polarization angles $\chi$, as indicated in Fig.~\ref{fig1}.

Instrumental broadening effects on the interference profiles, as illustratively shown in Fig.~\ref{fig2}(a), can reduce accuracy when combined with the uncertainty of the $R$ parameter, which is in fact the major source of inaccuracy, as demonstrated in Fig.~\ref{fig2}(b). The $E_0(\delta_{\rm T})$ curves in Fig.~\ref{fig2}(b) is just showing that it is not possible to extract an accurate value of $E_0(\delta_{\rm T})$ from a single $\varphi$-scan when $R$ is unknown.
\begin{figure*}
\includegraphics[width=3.2in]{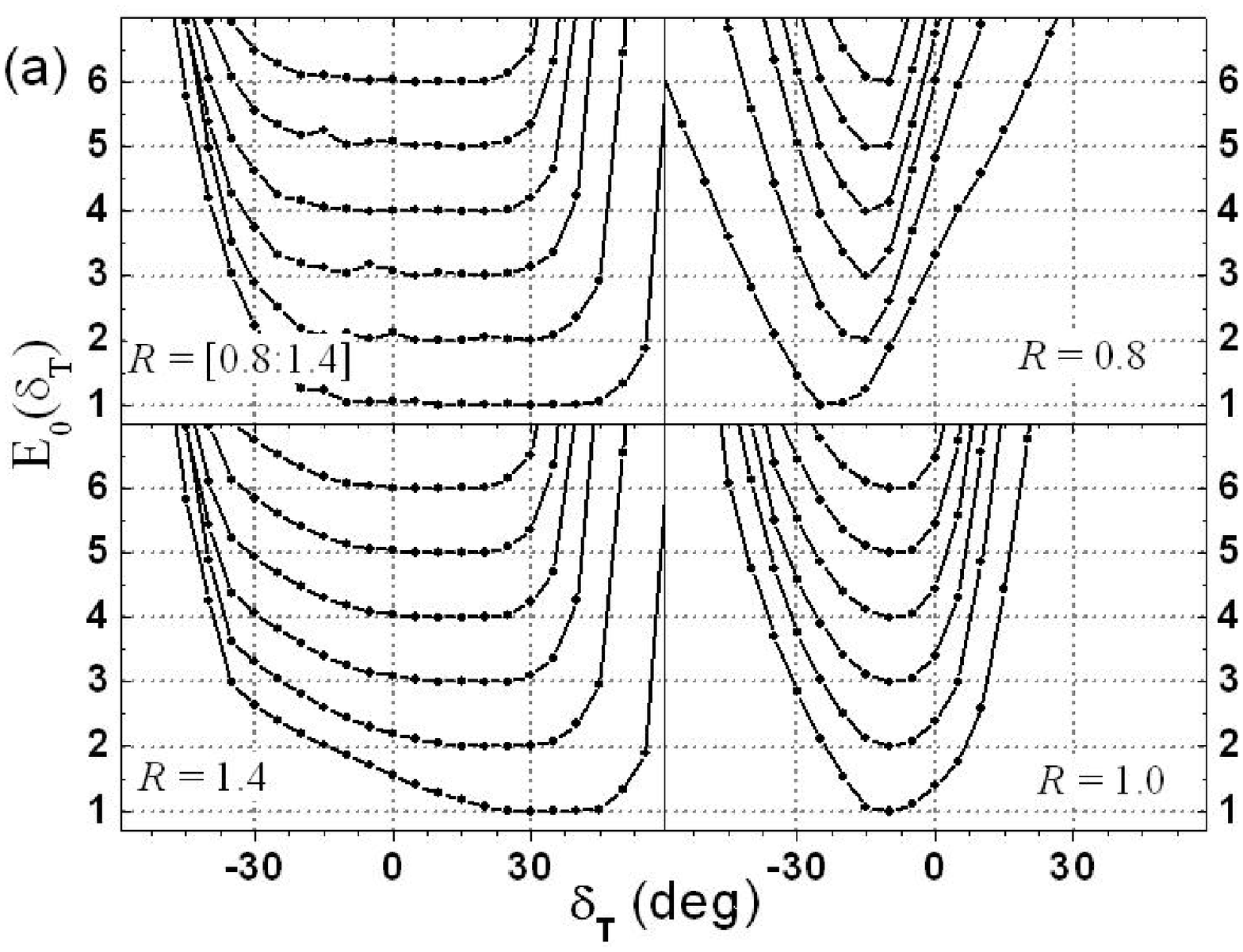}
\includegraphics[width=3.2in]{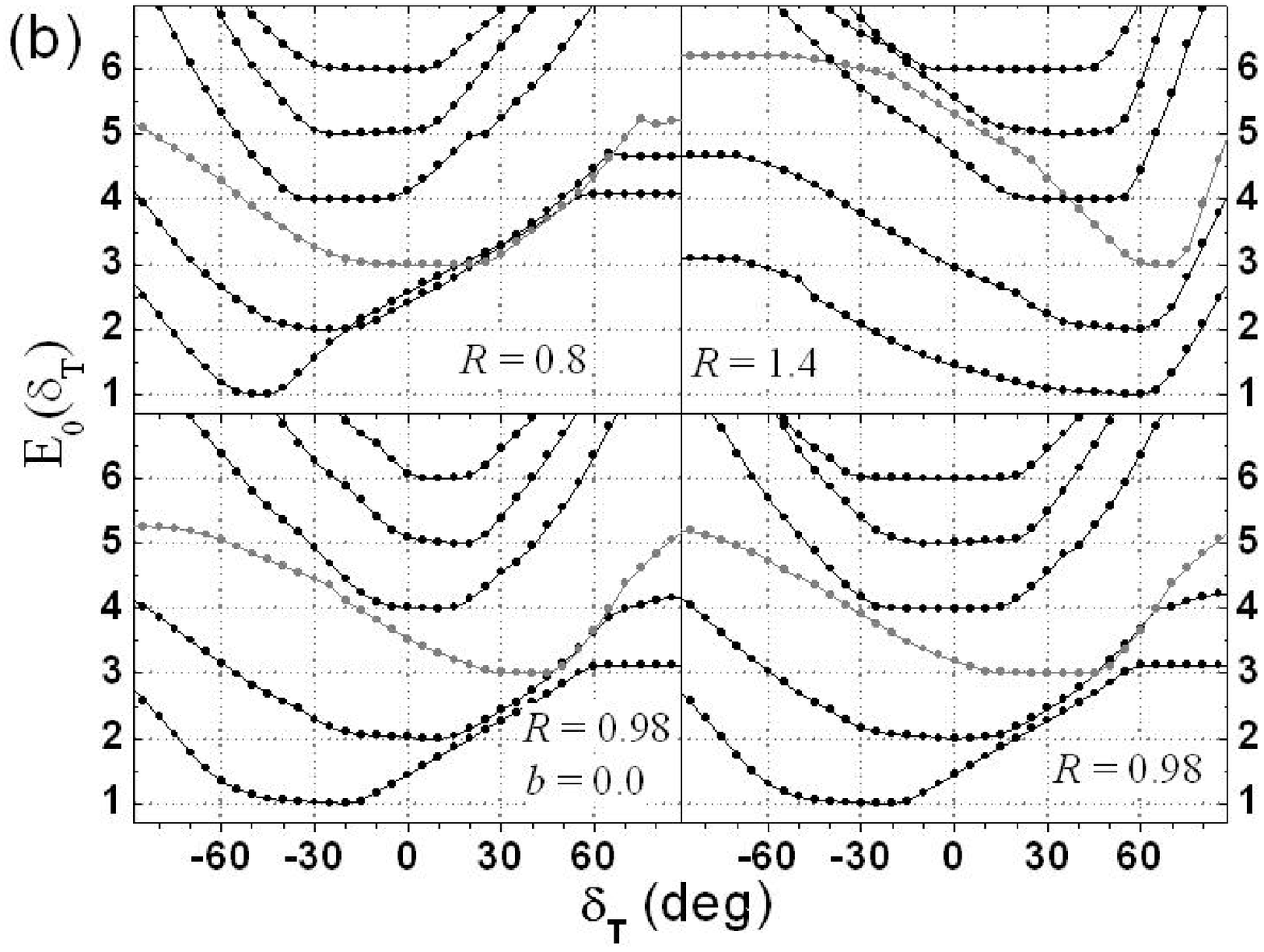}
\caption{\label{fig3} Absolute-mean deviation as a function of $\delta_{\rm T}$, $E_0(\delta_{\rm T})$, obtained for (a) the simulated scans and (b) the experimental scans in Fig.~\ref{fig1}. All curves are normalized by its minimum value and add to an integer for better visualization. The curves with minima equal to 1, 2, 3, 4, 5, and 6 correspond to those scans with $\chi = 8^{\circ}$, $12^{\circ}$, $16^{\circ}$, $20^{\circ}$, $24^{\circ}$, and $32^{\circ}$, respectively. Allowed range is $\bm{p} = [0.0010^{\circ}:0.0014^{\circ},\> R,\> 0.0:1.0,\> 0.0:3\bm{v}^2,\> \varphi_0\pm0.012^{\circ},\> 0.001^{\circ} : 0.006^{\circ}]$ where the $R$ values or ranges are shown in the figure for each case, and $\bm{v}^2$ changes the upper limit of the $b$ range with the polarization angle; here $\bm{v}^2=\sin^2\chi$.}
\end{figure*}

The best strategy, that we could elaborate, for accurate determination of triplet phases is composing polarization-dependent sets of azimuthal scans, as the one in Fig.~\ref{fig1}, and then, search for the value of $R$ that provides $\partial E_0 / \partial \delta_{\rm T} = 0$ as close as possible of a same $\delta_{\rm T}$ value. Here this search strategy has been applied in two sets of azimuthal scans: a simulated one that is free of instrumental effects such as statistic noise and sample misalignments, and another that is the experimental data in Fig.~\ref{fig1}. The $E_0(\delta_{\rm T})$ curves of the simulated $\varphi$-scans for several values of $R$ are shown in Fig.~\ref{fig3}(a) while Fig.~\ref{fig3}(b) shows the respective $E_0(\delta_{\rm T})$ curves for the experimental data.

\section{Conclusions}

The data analyses presented here have demonstrated that systematic and reliable phasing procedures are feasible. However, accuracy can be improved by optimizing the incident X-ray beam optics regarding energy resolution and angular divergences, mainly in the horizontal plane. A good instrumental precision is also required as well as low noise in the intensity data.

\begin{acknowledgments}
This work has been supported by the Brazilian Synchrotron Light Source (LNLS) under proposal No. D12A - XRD1 - 1264, FAPESP (proc. No. 02/10387-5), and CNPq (proc. No. 301617/95-3 and 150144/03-2). 
\end{acknowledgments}

\end{document}